# Kinetics of island growth in the framework of "planar diffusion zones" and "3D nucleation and growth" models for electrodeposition

S. Politi and M. Tomellini[(*)]

Dipartimento di Scienze e Tecnologie Chimiche, Università degli studi di Roma Tor Vergata,

Via della Ricerca Scientifica 00133 Roma (Italy)

**Abstract**

In the electrochemical deposition of thin films the measurement of the current-time curve does not allow for a direct determination of the nucleus growth law, electrode surface coverage and mean film thickness. In this work we present a theoretical approach suitable to gain insight into these quantities from the knowledge of nucleation density, solution parameters and current-time behavior. The model applies to both isotropic and anisotropic growth rates of nuclei and a study on the effect of nucleus shape and aspect ratio on the kinetics is presented. Experimental results from literature are also discussed in the framework of the present approach.

# 1-Introduction

In recent decades the importance of electrodeposition in Materials Science has been widely recognized in connection with the development of nanostructured materials for nanotechnology application [1]. Electrochemical deposition is an inexpensive, flexible and versatile technique suitable for the synthesis of novel nanostructures made up of metals, oxides and polymers. The mechanism of deposition is important since it determines, in large part, the structure and properties of the materials. In this context deposition ruled by nucleation and growth processes are winning attention in the literature also in view of the possibility to control both dimension and number of nuclei [2,3]. With the advent of the nanotechnology and the increasing demand to fabricate materials at the nanometer scale, it is relevant to advance in the understanding of this mechanism. This needs insight into island growth law, film thickness and surface coverage of the electrode.

The potentiostatic transient measurement is a suitable method for studying the initial kinetics of deposition process which has been attracting interest from both theoretical and experimental points of view [2–5].Theoretical models have been proposed for describing the current density in terms of either 3D- or 2D-nucleation and growth processes which have been profitably employed to interpret experimental chronoamperometric transients [4,5]. These methods can be classified within two main categories namely, those based on the "3D nucleation and growth" theory [6-7] and those exploiting the analogy with a planar diffusion process [8,9]. The latter approach has been mainly employing by researchers for analyzing experimental data at different levels of comprehension. This is due to the fact that the theory by Scharifker and Hills provides analytical kinetics, usually expressed in terms of normalized time and current, suitable for treating chronoamperometric curves. In this context most of the investigations deal with metal electrodeposition onto different substrates. Studies on the electrodeposition kinetics on various substrates such as metals [10,11], highly ordered pyrolytic graphite (HOPG) [12,13], vitreous carbon [14], carbon nanotubes fibers [15] and inside the nanosized pores of polyethylene membrane [16] have been performed, in order to investigate the effect of the substrate on both deposition process and final structure of the film. During the last years the latter model has also been applied to study electrodeposition of oxide [17] and polymeric films [18,19].

On the other hand, the former method, based on a detailed analysis of the nucleation and growth phenomena, provides transient kinetics in terms of integral expressions where, depending on nucleation and growth laws, the integrals do not always admit analytical solutions. Besides, in this method the knowledge of the nucleus growth law is required, an information not often available. These peculiarities of the model are the main limits to its applicability for describing experimental kinetics. Nevertheless, as discussed throughout the present paper, the "3D nucleation and growth" approach is, from the theoretical point of view, rigorous and quite exhaustive. In fact, the important physical processes occurring during film formation are properly considered in the model, where the probability theory is employed in order to determine the kinetics.

In this work we present a novel study by employing these two approaches to the model case of site-saturation nucleation, where all nuclei start growing at the same time. Although this nucleation mode is a limiting case, it is important and worthy to be studied for the following reasons: i) the distribution of nuclei can be assumed to be strictly random on the surface and ii) phantom nuclei do not enter the mathematical formulation of the kinetics. In addition, site-saturation nucleation (also named simultaneous nucleation) has been found to occur in electrodeposition [20–23]. According to point i) impingement among islands cannot be neglected and must be duly taken into account in the formulation of the model. According to point ii) phantom overgrowth, which is related to diffusion-type growths, does not take place and the Kolmogorov-Johnson-Mehl-Avrami (KJMA) theory holds true for describing phase transformation kinetics ruled by nucleation and growth. Detailed discussion on the concept of phantom nuclei and phantom overgrowth in KJMA model is reported in refs [ 24–27].

In the non-simultaneous nucleation, owing to the diffusional fields, nuclei are not randomly distributed on the electrode surface. Under this circumstance due to spatial correlation effects, nuclei are more dispersed on the surface and impingement events play a minor role when compared to the random case [28]. Monte Carlo simulations of planar diffusion zones have been performed for non-simultaneous nucleation where the size of diffusion zones depends upon on nucleus age [29]. In this paper an alternative to the mean-field approximation for the treatment of the overlap of planar diffusion zones has been proposed. It is found that for progressive nucleation the analytical model developed in ref. [30] is in accord with the computer simulations.

The purpose of this work is twofold, firstly to review the aforementioned models in their original formulation. Secondly, exploiting the two descriptions, to gain insight into the growth law of the islands and mean thickness of the film. It is shown that the two approaches are complementary in providing a more complete description of the electrodeposition kinetics.

The paper is organized as follows. In the first section the main equations of planar diffusion model are briefly reviewed. The second section is devoted to the "3D nucleation and growth" model where the critical region method is employed to determine the kinetics. The comparison between the two approaches is presented in the third section where growth related quantities are obtained by using both formulations. In the "Discussion" section these results are analyzed in the light of experimental data and simulations from the literature.

## 2- Results

### 2.1-"Planar diffusion zones" model

For diffusion-controlled mass transfer, the growth rate of a single hemispherical nucleus (mol s$^{-1}$) starting growing at time $t$=0 is given by [31]

$$I(t) = \pi^{\frac{1}{2}} D^{\frac{3}{2}} c \left( \frac{8\pi Mc}{\rho} \right)^{\frac{1}{2}} t^{1/2}, \tag{1a}$$

where $D$ is the diffusion coefficient of the species to be deposited, $c$ its bulk concentration, $\rho$ and $M$ density and molecular weight of the deposit, respectively. Using eqn.1a in the expression $I(t) = 2\pi \frac{\rho}{M} R_0^2 \frac{dR_0}{dt}$, $R_0$ being nucleus radius, the growth law is obtained as

$$R_0(t) = \sqrt{\beta_0 t}, \tag{1b}$$

with $\beta_0 = \left( \frac{2cMD}{\rho} \right)$.

A planar diffusion zone is defined by the equivalent area required to sustain the growth of an isolated hemispherical nucleus by linear diffusion (Fig.1a). The flux density at a planar surface reads [32],

$$J_p(t) = \frac{Dc}{\sqrt{\pi Dt}}, \tag{2a}$$

from which the equivalent area is obtained by imposing the identity

$$I(t) = J_p(t)\pi r_d^2(t) \tag{3a}$$

leading to

$$r_d(t) = \sqrt{\gamma t}, \tag{3b}$$

with $\gamma = D\left(\frac{8\pi Mc}{\rho}\right)^{1/2}$ greater than $\beta_0$. We point out that, by definition, eqns.3a-3b provide the exact $I(t)$ kinetics for the growth of an isolated hemispherical nucleus. In fact, although the actual diffusion process is spherical (and not planar), at mathematical level eqn.1a can be factorized in the product given by eqn.3a with a suitable choice of the $r_d$ function (eqn.3b). In the literature the disk of radius $r_d$ is called "planar diffusion zone" [32]. It follows that for an ensemble of $N_0$ (cm$^{-2}$) isolated nuclei, starting growing at $t$=0, the whole current density, $J(t) = zFN_0 I(t)$ (A cm$^{-2}$), becomes $J(t) = zFJ_p(t)N_0\pi r_d^2(t)$, namely

$$J(t) = zFDc\frac{\theta(t)}{\bar{\delta}(t)}, \tag{4}$$

where $\theta(t) = N_0\pi r_d^2(t)$ is the fraction of electrode surface covered by $N_0$ disks of radius $r_d$ and $\bar{\delta}(t) = \sqrt{\pi Dt}$ the diffusion length. Eqn.4 is the exact solution of the growth problem under consideration. In addition, since in the early stage of the deposition nuclei are well separated, eqn.4 provides the correct expression of the current density in this case as well.

In the Scharifker and Hills approach [9] the current density in the entere regime of growth is still computed through eqn.4 by identifying $\theta(t)$ with the coverage of $N_0$ disks of radius $r_d$ randomly distributed on the surface. This is done by using the KJMA model for phase transformations in 2D, which allows one to describe the whole area of the disks by taking into account the collision (impingement) among them [24–27]:

$$\theta(t) = 1 - \exp(-\theta_{ex}(t)) \tag{5a}$$

where

$$\theta_{ex}(t) = N_0 \pi r_d^2(t) \tag{5b}$$

is the extended surface. One notices that eqns.4-5 provide the correct long- and short- time behaviors of the current density. Extension of the exact solution for isolated nuclei to the entire regime of growth, by means of eqn.5a, is a mathematical conjecture whose validity has been subjected to experimental verification [4, 30]. The fact that $r_d > R_0$ has also been interpreted, in progressive nucleation, as the presence of a spatial region around central nucleus where further nucleation is hindered, an hypothesis supported by experimental data [4, 28]. Besides, since $\theta(t) < \theta_{ex}(t)$, the whole deposition rate is lower than that of $N_0$ isolated nuclei and the growth rate of a nucleus is expected to be lower than eqn.1, on average. As discussed below, this outcome can be tested through experiments.

On theoretical side, diffusional growth of an assembly of nuclei has been studied in detail in ref.[33] by solving a non-homogeneous diffusion equation with point-sinks. The solution is obtained in terms of Green functions which propagate the interactions among diffusion sinks (nuclei). A closed form of the solution is attained by exploiting a mean field approximation with concentration profile $c(z, t)$ that is a function of the coordinate along surface normal, $z$. In other words, correlation with neighboring nuclei leads to a symmetry breaking of the diffusion process, from spherical to planar. Noteworthy, it is shown in ref.[9] that Scharifker and Hills model (eqns.4,5) is in very good agreement with the mean-field solution and in better agreement with a more robust numerical simulation of the diffusional growth.

For the sake of completeness, it is worth recalling that Scharifker and Hills method has been applied to non-simultaneous nucleation as well. A series of works [30-32, 34-35] provide detailed studies on the determination of the $\bar{\delta}(t)$ function for non-simultaneous nucleation.

The planar diffusion zones approach was introduced to obtain an approximate solution of a nucleation and diffusional-growth problem that cannot be solved analytically. It is a useful phenomenological model which allows the transition from hemispherical (short-time) to planar (long-time) regimes of growth to be modeled in a quite simple fashion.

Eqns.4-5 eventually provide

$$J(t) = zFJ_p(t)\theta(t) = zFcD^{1/2}\frac{1-e^{-\pi N_0\gamma t}}{\sqrt{\pi t}}. \tag{6}$$

The volume of the deposit per unit of electrode surface, $V(t)$, can be determined from current density through the relationship $\frac{J(t)}{zF} = \frac{\rho}{M}\frac{dV(t)}{dt}$, as

$$V(t) = \frac{cMD^{1/2}}{\rho\pi^{1/2}}\int_0^t \frac{\left(1-e^{-\pi N_0\gamma t'}\right)}{\sqrt{t'}}\,dt'. \tag{7a}$$

By using the definition of $\beta_0$ and $\gamma$, one obtains $\frac{cMD^{1/2}}{\rho\pi^{1/2}} = \frac{\beta_0^{3/2}}{\gamma}$. A change of variable from $t'$ to $\eta^2 = \frac{t'}{t}$ in eqn.7a provides

$$V(t) = \frac{2}{\nu}\sqrt{\beta_0 t}\int_0^1 \left(1-e^{-S_{ex}(t)\nu\eta^2}\right)d\eta, \tag{7b}$$

where $S_{ex}(t) = \pi N_0\beta_0 t$ and $\nu = \frac{\gamma}{\beta_0} = \frac{\pi N_0 r_d^2}{\pi N_0 R_0^2}$ is the ratio between the (extended) areas of the diffusion zones and electrode/deposit interface ($\nu > 1$). It is worth noting that $\nu$ depends on deposit and solution concentration according to $v = \left(\frac{2\pi\rho}{Mc}\right)^{1/2}$, which is independent of diffusion coefficient of the species in solution. In eqn.7b only the $\beta_0$ parameter is linked to the diffusion coefficient. $S_{ex}(t)$ is the extended surface which enters the KJMA kinetics for the formation of the electrode/deposit interface. Since both $N_0$ and $\beta_0$ are constants, $S_{ex}(t)$ is in fact a reduced time. In the framework of both mean field approach of ref. [33] and eqn.7b, the reduced time $S_{ex}(t)v$ is the time after which interaction between nuclei starts to play a role.

*2.2-"3D nucleation and growth" model*

In this approach the formation of the deposit is described through nucleation on the electrode surface followed by the growth along the surface normal [36]. Nuclei are randomly distributed on the surface and both nucleation and growth laws are assumed to be known. The problem of finding the kinetics of film growth has been solved in ref.[6] by exploiting the KJMA model for a set of 2D layers randomly stacked along the surface normal. A different approach, based on the concept of "critical region", has also been proposed more recently in [37] leading to the same result. Here, we discuss the essential ideas of the latter method which has been employed to solve several problems in the field of phase transformations at the solid state [38–40]. In this approach one considers a generic point of the space, say Q, located at height $h$ from the surface and determines the probability that this point is not transformed by the new phase at time $t$ (Fig.1c). For a random distribution of nuclei this probability reads,

$$P_0(h, t) = \exp[-\xi_c(h, t)], \tag{8}$$

where $\xi_c(h, t)$ is proportional to the area of the region (critical region) in which nucleation events are capable of transforming the point Q up to time $t$ (Fig.1c). For hemispherical nuclei it is equal to

$$\xi_c(h, t) = H(R(t, 0) - h) \int_0^{t'_{mx}} \pi[R(t, t')^2 - h^2]\dot{N}(t')dt', \tag{9a}$$

where $H(x)$ is the Heaviside function ($H(x) = 1$, for $x > 0$; $H(x) = 0$, for $x < 0$), $R(t, t')$ the nucleus radius and $t'_{mx}$ satisfies the equation $R(t, t'_{mx}) = h$. In eqn.9a, $R(t, 0)$ is the maximum radius of the population of nuclei.

Application of eqn.9a to site-saturation nucleation and diffusional growth (eqn.1b), gives rise to

$$\xi_c(h, t) = \pi N_0[\beta_0 t - h^2]H\left(t - \frac{h^2}{\beta_0}\right).$$ 
(9b)

The transformed volume fraction per unit of electrode area is eventually computed through integration over $h$: $V(t) = \int_0^{R_0(t)} (1 - e^{-\pi N_0(\beta_0 t - h^2)})\ dh$. Changing variable from $h$ to $\eta = \frac{h}{R_0(t)}$, with $R_0(t) = \sqrt{\beta_0 t}$, eqn.9b eventually provides

$$V(t) = \sqrt{\beta_0 t} \int_0^1 \left[1 - e^{-S_{ex}(t)(1 - \eta^2)}\right] d\eta.$$ 
(10)

It is possible to show that at long-time the current density attained from eqn.10 does not scale as the Cottrell equation [41]. Such a behavior limits the applicability of eqn.10 for describing experimental data.

*2.3 Kinetics of island growth*

By comparing eqn.10 with eqn.7b it follows that the "planar diffusion zones" and "3D nucleation and growth" models, give quite different kinetics. This is shown in Fig.2 where the kinetics (normalized to $\sqrt{\beta_0 t}$) are reported as a function of $S_{ex}(t)$ for several values of $\nu$. In the same figure it is also shown the behavior of the normalized current density ($i = J/J_{max}$). The curves coincide only at the very beginning of the deposition. In fact, for $S_{ex}(t) \to 0$ a series expansion of eqns.7b and 10 gives, respectively, $\frac{V(t)}{\sqrt{\beta_0 t}} = 2S_{ex}(t) \int_0^1 \eta^2 d\eta = \frac{2}{3} S_{ex}(t)$ and $\frac{V(t)}{\sqrt{\beta_0 t}} = S_{ex}(t) \int_0^1 (1 - \eta^2) \, d\eta = \frac{2}{3} S_{ex}(t)$. This limiting case provides the growth law $R \propto \sqrt{t}$, as expected in the early stage of deposition.

The reason of the discrepancy between the two approaches is due to the $R_0(t)$ function used in the 3D nucleation and growth model, that is the growth law of an isolated hemispherical nucleus. The effect of interactions of diffusional fields on nucleus growth is not taken into account by eqn.10. On the contrary, this effect is included in the planar diffusion model through an appropriate description of the impingement among diffusion disks. Since eqns.6 is obtained in the framework of a mean field approach, that is the KJMA model, it implies $\frac{J(t)}{zF} = J_p(t) N_0 \bar{s}_1(t)$, where $\bar{s}_1$ is the mean area of planar diffusion zone. It follows that the mean current to a nucleus is $\bar{I}_1(t) = J_p(t) \bar{s}_1(t)$. The growth of the mean nucleus is linked to planar diffusion across a surface of area $\bar{s}_1(t)$. As far as nucleation in the actual and equivalent planar systems are concerned, they are realization of the same Poisson process, with density of dots equal to $N_0$. On this basis it is possible to consider the same random arrangement of dots for both systems. In other words, each diffusion disk is centered on a nucleus. Such a representation can be useful to illustrate the effect of impingement among disks on nucleus growth. Let us consider the case of binary collisions between diffusion disks. For isolated couples of nuclei (of radius $R$) at relative distance $\zeta$, the surface of each nucleus is $\sigma_1 = \pi R^2 \left(1 + \frac{\zeta}{2R}\right) H(2R - \zeta) + 2\pi R^2 H(\zeta - 2R)$. The area of the planar diffusion zone, per nucleus, is given by $s_1 = \pi r_d^2 - \frac{1}{2} \omega(r_d, \zeta) H(2r_d - \zeta)$ with $\omega(r_d, \zeta) = 2r_d^2 \cos^{-1} \frac{\zeta}{2r_d} - \frac{\zeta}{2} \sqrt{4r_d^2 - \zeta^2}$ area of the overlap between disks and $r_d(t) = \sqrt{\gamma t}$ (eqn.3b). On the same token, let us consider a configuration of disks subjected to two symmetric binary collisions (respect to the circle center), and nuclei at relative distance $\zeta$ which can underwent at most to two binary collisions with neighboring nuclei. In this case the area of the diffusion zone is $s_2 = \pi r_d^2 - \omega(r_d, \zeta) H(2r_d - \zeta)$ and the surface of each nucleus is $\sigma_2 = 2\pi R^2 \left(\frac{\zeta}{2R} H(2R - \zeta) + H(\zeta - 2R)\right)$. For these configurations the growth rate is determined from the relation $I_1(t) = J_p s_n$, that is $\frac{\rho}{M} \sigma_n \frac{dR}{dt} = J_p s_n$ (with $n$=1,2), according to

$$\sigma_n(R) \frac{dR}{dt} = \frac{Mc(D)^{1/2}}{\rho \sqrt{\pi t}} \left[\pi r_d^2 - \frac{n}{2} \omega(r_d, \zeta) H(2r_d - \zeta)\right]. \tag{11a}$$

Specifying the functional form of the various terms of eqn.11a, the differential equation becomes

$$\frac{dw}{d\tau} = \frac{3}{2}\sqrt{\tau}\frac{\left[1 - \frac{n}{\pi}\left(\cos^{-1}\frac{1}{2\sqrt{\nu\tau}} - \frac{1}{2\sqrt{\nu\tau}}\sqrt{1 - \frac{1}{4\nu\tau}}\right)H(4\nu\tau - 1)\right]}{\left[\left(\frac{2-n}{2} + \frac{n}{4\sqrt[3]{w}}\right)H(8w - 1) + H(1 - 8w)\right]},$$  (11b)

where $w = \left(\frac{R}{\zeta}\right)^3$ and $\tau = \frac{\beta_0 t}{\zeta^2}$.

Eqn.11b shows that in the case of well separated planar diffusion zones, $2R < 2r_d < \zeta$, i.e. $4\nu\tau < 1$ and $8w < 1$, the growth law $R(t) = R_0(t) = \sqrt{\beta_0 t}$ is obtained. When overlap between diffusion disks takes place, the $R(t)$ function is expected to deviate from the parabolic law owing to the second term in the square brackets of eqn.11a. According to eqn.11b the displacement increases with $\nu$. In fact, the maximum value of the last term in the brackets of eqn.11b is attained for $\nu\tau \gg 1$ and it is equal to $\frac{n}{2}$. Numerical solutions of eqn.11b are displayed in Fig.3 for the $\tau$-interval in which nucleus growth is unimpeded by collisions with the two neighboring nuclei (n=2). This computation indicates that deviation of the solution (eqn.11b) from parabolic growth is larger the higher $n$ and $\nu$.

On the basis of this argument we determine the nucleus growth law, in the "3D nucleation and growth" model, which makes eqn.7b coincident with eqn.10. To this purpose, we employ the more general expression $R(t) = \sqrt{\beta(t)t}$, for which the solution of the kinetics, eqn.10, still holds provided $\beta(t)$, in place of $\beta_0$, is used in the expression of the extended surface. Since for $t \to 0$ the two kinetics coincide, we set $\beta(0) = \beta_0$. In terms of the quantity $x(t) = \frac{\beta_0}{\beta(t)}$, eqn.10 becomes $V(t) = \frac{\sqrt{\beta_0 t}}{\sqrt{x(t)}}\int_0^1\left[1 - e^{-\frac{S_{ex}(t)}{x(t)}(1-\eta^2)}\right]d\eta$. The equality between this relationship and eqn.7b eventually provides

$$\frac{\nu}{2\sqrt{x(t)}}\int_0^1\left(1 - e^{-\frac{S_{ex}(t)}{x(t)}(1-\eta^2)}\right)d\eta = \int_0^1\left(1 - e^{-S_{ex}(t)\nu\eta^2}\right)d\eta,$$  (12)

that is an equation for $x(t)$.

Eqn.12 has been solved, numerically, to obtain the actual growth law of nuclei, $R(t) = \sqrt{\frac{\beta_0 t}{x(t)}}$ and the fraction of electrode surface covered by islands, $S(t) = 1 - e^{-\frac{S_{ex}(t)}{x(t)}}$. In "3D nucleation and growth" statistical approach, growth law is the same for all nuclei. Accordingly, the extended radius $R(t)$ has to be considered an average over spatial configurations of nuclei. The outcomes of the computation are reported in figs.4 for the quantity $\tilde{R}(t) = \sqrt{\pi N_0} R(t) = \sqrt{\frac{S_{ex}(t)}{x(t)}}$. Fig.4 shows that the parabolic growth, employed in the original formulation of the model, is satisfied at the very beginning of deposition where $\tilde{R}(t) \rightarrow \tilde{R}_0(t) = \sqrt{\pi N_0 \beta_0 t} = \sqrt{S_{ex}(t)}$. Moreover, at long time eqn.12 provides $x(t) = \frac{\nu^2}{4}$, i.e. $R(t) = \frac{2\sqrt{\beta_0 t}}{\nu}$, which satisfies Cottrell equation.

The evolution of the normalized radius, $\tilde{R}$, is in accord with the power law $\tilde{R} = a S_{ex}^{\kappa}$ as indicated by the good fit to the numerical solutions reported in Fig.4 for the early stage of growth (coverage range $0 < S < 0.15$). This implies the scaling $\tilde{R}(t) = bt^{\kappa}$, with $b = a(\pi N_0 \beta_0)^k$. The behavior of both power exponent ($\kappa$) and pre-factor ($a$) with $\nu = \frac{\gamma}{\beta_0}$ is displayed in Fig.5, where the power exponent is in the interval $0.19 < \kappa < 0.5$, for $2 < \nu < 130$. These results can be discussed in connection with the growth law obtained by the binary collision model (eqn.11b). In fact, in eqns.11-12, $\zeta$ can be taken of the order of magnitude of the mean-distance among nuclei, $\bar{\zeta}$, that is, for a Poissonian distribution, $\bar{\zeta} = \frac{1}{2\sqrt{N_0}}$. Therefore, in eqn.11b $\tau = \frac{\beta_0 t}{\bar{\zeta}^2} \approx S_{ex}(t)$ and $\left(\frac{R(t)}{\bar{\zeta}}\right) \approx \sqrt{\pi N_0} R(t) = \tilde{R}(t)$, i.e. the kinetics displayed in Figs.3,4 are expressed in terms of similar variables. It is found that also the solutions of eqn.11b (Fig.3) are in accord with the power function. The computation provides a nearly constant value $\kappa_{n=1} \cong 1/2$ and a $\kappa_{n=2}$ which is a decreasing function of $\nu$. The $\kappa_n$ exponents are also reported in Fig.5 and are found to be greater than those obtained from eqn.12. This entails a non-negligible effect of higher order collisions among diffusion disks on growth law. A rough estimate of the lower-bound value of $\kappa$ can be obtained by considering isolated growing nuclei and overlapping diffusion disks according to the argument that follows. Let us singled out a diffusion disk whose growth is hindered by collisions with neighboring disks. In this case, the diffusion zone of the considered disk is the Voronoi cell of the nucleus located in its center [42–44]. Denoting with $\bar{A} = \frac{1}{N_0}$ the mean area of the Voronoi cells, in the case of isolated nuclei eqn.11a provides $2\pi R^2 \frac{dR}{dt} = \frac{Mc(D)^{1/2}}{\rho\sqrt{\pi t}}\bar{A}$ that becomes, in terms of $w$ and $\tau$ variables, $2\pi v \frac{dw}{d\tau} = \frac{3}{\sqrt{\tau}}$. This equation leads to the scaling $R(\tau) \propto \tau^{1/6}$, that is $\kappa \cong 0.17$. This limiting case is expected to occur for large values of $v$ being this quantity proportional to $\gamma$. Noteworthy, the $\kappa$ exponents computed from eqn.12 are found in agreement with this asymptotic behavior (dashed horizontal line in Fig.5).

As far as the $v$ values are concerned, in real systems they are usually larger than $v=50$, and this is due to the quite large ratio between concentration of the species in the deposit and in solution [20–23]. Such large values of $v$ entail low rates of the substrate coverage ($\dot{S}(t)$), when compared to the parabolic growth ($\kappa = \frac{1}{2}$), as Fig.6 shows for $v = 70, 100$ and $130$. For parabolic growth the behavior of $S(t)$ vs $S_{ex}(t)$ is independent of $v$ (solid gray line in Fig.6).

### 2.4. Anisotropic nucleus growth

The approach developed above for hemispherical nuclei can be extended to deal with less symmetric shapes resulting from anisotropic growth. In this section we study the case of ellipsoidal nuclei where the semi-axes on the surface plane are assumed to be equal. To model diffusional growth, the time dependence of the semi-axes is taken to be proportional to the square root of time. Equation for the nucleus surface is $\frac{x^2+y^2}{A^2 t} + \frac{z^2}{B^2 t} = 1$ where $(z > 0)$ and $(x, y)$ are the coordinates normal to the electrode surface and on surface plane, respectively. The aspect ratio of the nucleus ($\alpha$) is here defined as the ratio between the two semi-axis: $\alpha = \frac{B}{A}$.

In the computation that follows the evolution of the volume of the deposit is considered to be given by eqn.7b for ellipsoid as well. This assumption is supported by the results of refs.[45,46] which shows that the difference between the rate of diffusion to an ellipsoid and to an equivalent hemisphere is small for $0.2 < \alpha < 1$. Under this circumstance, $A$ and $B$ quantities are estimated, as a function of the aspect ratio, from eqn.1a, namely $I \equiv \frac{\rho}{M}\frac{dv}{dt} = \pi \frac{\rho}{M}\beta_0^{3/2}t^{1/2}$, v being nucleus volume. For ellipsoid $v = \frac{2\pi}{3}A^3 t^{3/2}\alpha$ and one obtains

$$A = \frac{\sqrt{\beta_0}}{\alpha^{1/3}}. \tag{13}$$

The 3D nucleation and growth model above presented can be generalized to anisotropic growth of the nucleus. It is possible to show (see the Appendix for details) that in this case

$$\xi_c(h,t) = \pi N_0 A^2 t \left(1 - \frac{h^2}{B^2 t}\right). \tag{14}$$

Inserting eqn.14 in eqn.8, and following the same computation pathway as in sect.2.2, the volume of the deposit reads

$$V(t) = \alpha A t^{1/2} \int_0^1 \left[1 - e^{-\pi N_0 A^2 t (1-\eta^2)}\right] d\eta \tag{15a}$$

that becomes, using eqn.13,

$$V(t) = \alpha^{2/3} \sqrt{\beta_0 t} \int_0^1 \left[1 - e^{-\frac{S_{ex}(t)}{\alpha^{2/3}}(1-\eta^2)}\right] d\eta, \tag{15b}$$

with $S_{ex}(t) = \pi N_0 \beta_0 t$.

Eqn.15a is the generalization of eqn.10 in the case of parabolic growth, i.e. with constant $A$. As in section 2.2, the actual growth law of nuclei is obtained by comparing eqn.15b with eqn.7b:

$$\frac{\nu \alpha^{2/3}}{2\sqrt{x_\alpha(t)}} \int_0^1 \left[1 - e^{-\frac{S_{ex}(t)}{x_\alpha(t)\alpha^{2/3}}(1-\eta^2)}\right] d\eta = \int_0^1 \left(1 - e^{-S_{ex}(t)\nu\eta^2}\right) d\eta, \tag{16}$$

where $x_\alpha(t) = \left(\frac{A(0)}{A(t)}\right)^2 = \frac{\beta_0}{A(t)^2 \alpha^{2/3}}$.

By scaling the $\beta_0$ parameter according to $\beta_0 \to \beta_0/\alpha^{2/3}$, that is $\nu \to \nu\alpha^{2/3}$ and $S_{ex}(t) \to S_{ex}(t)/\alpha^{2/3}$, eqn.12 becomes equivalent to eqn.16. This implies the identity $x_\alpha(\nu, S_{ex}) = x\left(\nu\alpha^{2/3}, \frac{S_{ex}}{\alpha^{2/3}}\right)$, with $x \equiv x_{\alpha=1}$ defined by eqn.12. Consequently, the time dependence of the reduced semi-axis of the ellipsoid, $\tilde{R}_\alpha = A(t)\sqrt{\pi N_0 t} = \frac{1}{\alpha^{1/3}}\sqrt{\frac{S_{ex}(t)}{x_\alpha(t)}}$, matches the power law behavior previously obtained for hemispherical nuclei with scaled exponent, $\kappa(\nu\alpha^{2/3})$.

The numerical solution of eqn.16 allows us to estimate $S(t)$ and, with it, the mean thickness of the deposit, $\bar{h}(t) = \frac{V(t)}{S(t)}$, as

$$\sqrt{\pi N_0}\bar{h}_\alpha(t) = \frac{2}{\nu}\sqrt{S_{ex}(t)}\frac{\int_0^1 \left(1 - e^{-S_{ex}(t)\nu\eta^2}\right)d\eta}{1 - e^{-\frac{S_{ex}(t)}{\alpha^{2/3}x_\alpha(t)}}} \tag{17}$$

and

$$\frac{\bar{h}_\alpha(t)}{R_\alpha(t)} = \frac{2\alpha^{1/3}}{\nu}\sqrt{x_\alpha(t)}\frac{\int_0^1 \left(1 - e^{-S_{ex}(t)\nu\eta^2}\right)d\eta}{1 - e^{-\frac{S_{ex}(t)}{\alpha^{2/3}x_\alpha(t)}}} \quad . \tag{18}$$

The behavior of the reduced mean thickness as a function of time and aspect ratio is reported in Fig.7a for $\nu = 70$. In panel b the fraction of the substrate surface covered by the deposit is also shown as a function of the same variables. In the early stage of growth, $S_{ex}(t) \to 0$, and the series expansion of eqn.18 provides $\frac{\bar{h}(t)}{R_\alpha(t)} \to \frac{2}{3}\alpha$. For large $t$ eqn.16 provides $\frac{2}{\nu\alpha^{2/3}}\sqrt{x_\alpha(t)} \to 1$, that is $x_\alpha(t) \to \frac{\nu^2}{4}\alpha^{4/3}$ which entails the asymptotic behavior of the exponent $\kappa \to 1/2$. Therefore, parabolic growth is obeyed once the substrate is completely covered by the deposit.

Current transients for cup-shaped nuclei have been computed in ref.[47] on the basis of the planar diffusion zones approach. The dependence of $J$ with contact angle has been modeled and an experimental technique for measuring contact angle during electrodeposition proposed in [48]. As regard the present approach, for cup-shaped nuclei the equation for the growth law (eqn.12) should be modified according to ref.[47].

## 3- Discussion

In the previous sections, by employing "3D nucleation and growth" and Scharifker and Hills approaches, we determined the kinetics of island growth and electrode surface coverage. To this end, in the former model nucleus growth law has been considered as an unknown. This is due to fact that parabolic growth, used in the original formulation of the approach, is obeyed by isolated nuclei, i.e. in the initial stage of deposition. In addition, it may happens that the same equation for current density is obtained for both models but for different shapes of the nucleus, namely hemisphere and right-cone. This aspect has been stressed in ref.[7] where an equation mathematically equivalent to eqn.6 has been obtained by a 3D nucleation and growth model for right-circular cone nuclei. In fact, for diffusion controlled growth in both directions, the two-rate method developed in [7] provides (see also the Appendix)

$$J(t) = \frac{zF\rho}{M} \frac{\mu_2}{\sqrt{t}} (1 - e^{-4\pi N_0 \mu_1^2 t}).$$  (19)

where $p_r = \frac{\mu_1}{\sqrt{t}}$ and $p_h = \frac{\mu_2}{\sqrt{t}}$ are the growth rates parallel and normal to the substrate plane, respectively. The growth law of the cone basis is $R(t) = 2\mu_1\sqrt{t}$. It follows that, for $\mu_1 = \frac{\sqrt{\gamma}}{2}$ and $\mu_2 = \left(\frac{D}{\pi}\right)^{1/2} \frac{cM}{\rho}$ eqn.19 coincides with eqn.6. The argument raised in [7], points to the evidence that the two models are mathematically isomorphous, though physically different, and that an experimental distinction between the two seems difficult. In what follows we discuss how to deem which one of the two models is more suitable for describing current time transients.

Eqn.19 and eqn.6 can be written in the general form

$$J(X_e) = K \frac{1-e^{-X_e}}{\sqrt{X_e}},$$  (20)

where $K$ is a constant and $X_e$ is the argument of the exponential which depends, linearly, on time. Both $K$ and the linear dependence of $X_e$ on time, are model dependent. At maximum $\frac{dJ}{dt} = \left(\frac{dX_e}{dt}\right)\frac{dJ}{dX_e} = 0$, namely $\frac{dJ}{dX_e} = 0$ since $X_e$ is proportional to $t$. It follows that the maximum of $J$ occurs at $e^{-X_e^*}(1 + 2X_e^*) = 1$, that is $X_e^* = 1.256$. We point out that the term $(1 - e^{-X_e})$ in eqn.20 has a well defined physical meaning which is model dependent. In the "planar diffusion zones" model it is the total area of the planar zones; in the "3D nucleation and growth" approach it is the fraction of electrode surface which is covered by nuclei. Consequently, the maximum current implies a surface coverage equal to $1 - e^{-X_e^*} = 0.7$. The meaning of this surface coverage is model dependent:

i) *"Planar diffusion zones"* model: 0.7 is the surface fraction of electrode covered by planar diffusion zones that is different from the surface coverage by islands (according to section 2.3).

ii) *"3D nucleation and growth"* model: 0.7 is the fraction of the electrode surface covered by the deposit.

   Therefore, provided that eqn.20 well describes the experimental transients, an experimental analysis of surface coverage at maximum current could identify the model which is more suitable for the system under investigation. An example of this kind of analysis can be done using the experimental results on Cu nucleation [23]. In this paper the authors report both transient currents and Atomic Force Microscopy (AFM) images at overpotentials where diffusion is rate determining. In particular, as shown by the authors, the set of data at pH = 1 and $c = 0.01$ are well described by eqn.20 for the instantaneous nucleation mechanism (Fig.5 ref.[23]). For this set of transients the authors also report AFM images of the deposit at sampling time longer than the one at maximum (for example Fig.8c, 9a, 11a in [23]). Although in ref.[23] the fractional coverage has not been estimated, it is apparent that it is much lower than 0.7. This lends support to an interpretation of eqn.20 based on the "planar diffusion zones" approach. Under this assumption the extended surface in Eqn.20 is given by $X_e = \nu S_{ex}$. However, the coverage of electrode surface by the deposit is not accessible through eqn.21 alone; to this end we exploit the results of section 2.3. From ref.[23] we get (figs.8,9 in [23]) $N_0 \approx 9.2 \ 10^7 \text{cm}^{-2}$, $c \approx 0.01M$, $D \approx 8 \ 10^{-6} \text{cm}^2 \text{ s}^{-1}$ and $t = 3s$ from which one obtains $S_{ex} = \pi N_0 \beta_0 t \approx 1$. The fractional coverage is given by $S = 1 - e^{-\tilde{R}^2}$. Using the power law dependence $\tilde{R} = a S_{ex}^{\kappa}$ at the appropriate parameter values ($a \approx 0.27$ and $\kappa(\nu) \approx 1/6$, from Fig.5) we end up with a $S \approx 8\%$. This is a more reasonable figure when compared to the value $S \approx 70\%$ predicted by the "3D nucleation and growth" approach for right cone. Another study by the same authors on Cu deposition on glassy carbon in ammonical solutions has been presented in ref.[49]. The joint analysis of transient currents and film morphology by AFM allows one to estimate the surface coverage. For this system deposition at pH=4 occurs by simultaneous nucleation and the AFM images have been recorded after 10s deposition, i.e. well beyond current transient maxima (Fig.13 in [49]). For example, at $c \approx 0.005M$, $D \approx 7.7 \ 10^{-6} \text{cm}^2 \text{ s}^{-1}$ (measured by the Cottrell equation), $N_0 \approx 7 \ 10^8 \text{ cm}^{-2}$ and $t = 10s$ one obtains $S_{ex} = \pi N_0 \beta_0 t \approx 12$. Following the same computation as above we estimate a $S \approx 15\%$. From the data set of the same authors it follows that the typical nucleus size, at sampling time, is of the order of magnitude of 50nm [23]. For the parameter values above reported, this entails $R \ll \sqrt{\pi D t}$ i.e. $\bar{R}^2 \ll S_{ex}$. As a consequence, the linear-type contribution in the expression of the growth rate of isolated nucleus, can be neglected as generally assumed in the literature [31] and in eqn.1a as well. Results of the computation of the surface coverage for $v$-value compatible with the systems above discussed have been reported in fig.6 for two values of the aspect ratio. In the same figure a computer simulation of 3D KJMA-type phase transformation is reported showing the surface aspect at different $S$ [50].

Before concluding, we consider again the nucleus growth law. It is worth pointing out that the power exponents obtained in Fig.4, are representative of an average of the growth law over coverage range $0 < S < 0.15$. At the very beginning of the deposition the exponent is always equal to ½, independently of $v$, as above demonstrated. Variation of the exponent of the power law during islands growth has already been inferred in refs.[51, 52]. Computer simulations of island formation during electrodeposition do show that for diffusion limited growth of an isolated island the power exponent changes, at longer time, from ½ to 1/6 [52]. Such a transition has also been confirmed by experiments on individual nucleus growth by means of real-time kinetics [53]. The transition is found to occur when the substrate surface coverage is much less than unity [52]. These results, together with the non-Poissonian distribution of nuclei in progressive nucleation, give support to a description of the growth process based on the concept of "2D exclusion zone" for nucleation [52]. In the framework of a mean field theory, the growth law should be compliant with such a behavior of the power exponent. This aspect is not dealt with by the original formulation of "3D nucleation and growth" approaches where nucleus growth is parabolic (diffusional growth). In addition, for right-circular cone with non-parabolic growth the two-rate model implies $p_r \propto t^{k-1}$ ($k \neq 1/2$). In this case the functional form of the transient becomes $J \propto t^{k-1} \left(1 - e^{-\mu t^{2k}}\right)$, whose mathematical form differs from eqn.19. At long time it does not scale according to the Cottrell equation, an inconsistency that is in common with the transient modeling of hemispherical nuclei with parabolic growth, as discussed in ref.[41].

To summarize this section we notice that the "3D-nucleation and growth" model for right-cone with parabolic growth could be distinguished from the "planar diffusion zones" approach by means of surface coverage and current transient measurements. In the framework of the former model, an attempt to reduce the growth rate, at least "phenomenologically" by using a power exponent $k < 1/2$, leads to incorrect behavior of the current at long-time. On the other hand, in the approach here presented, the growth law is obtained by comparing the two approaches; in doing so the long-time limit is also satisfied.

**4- Conclusions**

In this work we have employed "*planar diffusion zones*" and "3*D nucleation and growth*" methods for modeling island growth in electrodeposition by site-saturation nucleation. Owing to its versatile analytical form, the former is mainly employed by researchers for analyzing chronoamperometric transients. In fact, it provides a good approximate solution of an involved diffusional-growth problem due to interference between diffusion fields. On the other hand, the latter exploits a stochastic approach for describing nucleation and growth on the electrode surface. In this case the kinetics is, in general, less versatile as it is given by integral expressions.

These methods are usually considered as alternative to each other. The present study shows that they are complementary for a more comprehensive description of the kinetics. Combination of both methods allows one to gain information on nucleus growth law, mean film thickness and surface coverage of the electrode, quantities that are not accessible using just one of these methods alone.

The main results of this study can be summarized as follows

- In the interval of coverage relevant for potentiostatic transients, the growth law of actual nucleus (extended nucleus) is in accord with a power law. Power exponent depends on deposit and solution parameters through $\nu$ and ranges between $\frac{1}{2} < \kappa(\nu) < \frac{1}{6}$.
- A differential equation for nucleus radius has been derived in the model case of binary collisions among diffusion disks. The solution is in accord with the scaling of nucleus size obtained through eqn.12.
- The comparative analysis is extended to deal with ellipsoidal nucleus. It is shown that power law holds in this case as well, with $\kappa(\nu)$ independent of aspect ratio ($\alpha > 0.2$). The fraction of electrode surface covered by islands is determined as a function of reduced time, $\nu$ parameter and aspect ratio.

**Acknowledgement**

The authors are indebted with Dr. E. Tamburri for the helpful discussions and the critical reading of the manuscript.

**Appendix**

Let Q be a point at height $h$ from the electrode surface and O its projection on the electrode surface. The area of the capture zone of point Q for nuclei born at time $t' < t$ is equal to $\pi r(t, t')^2$ where $r(t, t') \equiv \overline{ON}$ and the point $N$, on the electrode surface, is such that a nucleus born at $N$ will reach point Q at time $t$ (Fig.A1). Since the points on the nucleus surface satisfy the equation, $\frac{x^2 + y^2}{A^2(t-t')} + \frac{z^2}{B^2(t-t')} = 1$, we get

$$\frac{r(t,t')^2}{A^2(t-t')} + \frac{h^2}{B^2(t-t')} = 1, \tag{A1}$$

namely

$$r(t,t')^2 = A^2(t - t') - \frac{h^2}{\alpha^2}, \tag{A2}$$

where $\alpha = \frac{B}{A}$. Since $r(t,t')^2 > 0$ and $t' < t'_{mx} = t - \frac{h^2}{A^2\alpha^2} = t - \frac{h^2}{B^2}$, $\xi_c(h, t)$ is eventually computed by summing over the population of nuclei,

$$\xi_c(h,t) = H(t'_{mx})\pi \int_0^{t'_{mx}} r(t,t')^2 \, \dot{N}(t')dt' = H(t'_{mx})\pi \int_0^{t-\frac{h^2}{B^2}} (A^2(t-t') - \frac{h^2}{\alpha^2})\dot{N}(t')dt'. \tag{A3}$$

For site-saturation nucleation eqn.A3 gives eqn.14:

$$\xi_c(h,t) = \pi N_0 A^2 t \left(1 - \frac{h^2}{B^2 t}\right) H\left(1 - \frac{h^2}{B^2 t}\right). \tag{A4}$$

In what follows we illustrate the two-rate model of ref.[6]. In this method growth rates $p_r$ and $p_h$ are defined for "on-plane" and "normal" directions, respectively. These quantities provide the time dependence of the coordinates, $r$ and $h$, of a point lying on the nucleus surface at time $t$:

$$r(t,w) = \int_w^t p_r(\xi)d\xi \qquad (A5)$$

$$h(t',w) = \int_{t'}^w p_h(\xi)d\xi. \qquad (A6)$$

In these equations, $t'$ is the birth time of the nucleus, and $w > t'$ is the time of arrival of the nucleus at height $h$, i.e. the semi-axis of the ellipsoid at time $w$ equals $h$: $B^2(w-t') = h^2$. Eqns.A5-A6 provide $h = h(t',w)$ that can be inverted to get $w = w(t',h)$. Next, by considering the 2D phase transition on the layer at height $h$ we obtain (fig.A1): $\xi_c(h,t) = H(t'_{mx})\pi \int_0^{t'_{mx}} r(t,w(t',h))^2 \dot{N}(t')dt'$ with $w(t'_{mx},h) = t$, which becomes, using eqn. A5,

$$\xi_c(h,t) = H(t'_{mx})\pi \int_0^{t'_{mx}} \dot{N}(t') \left[ \int_{w(t',h)}^t p_r(\xi)d\xi \right]^2 dt', \qquad (A7)$$

that is the expression derived in ref.[6]. Next, we apply eqn.A7 to the ellipsoidal and cone-shaped nuclei in diffusional growth. To this purpose the $p_r$ and $p_h$ functions have to be determined.

*Ellipsoidal nuclei.* In this case $p_r$ and $p_h$ (eqn.A5-A6) should satisfy eqn.A1. It is possible to show that $p_r(t) = \frac{A}{2\sqrt{t-w}}$ and $p_h(t) = \frac{B}{2\sqrt{t-t'}}$ are the appropriate functions. Use of these rates in eqn.A7 provides

$$\xi_c(h,t) = H\left(t - \frac{h^2}{B^2}\right)\pi \int_0^{t-\frac{h^2}{B^2}} \dot{N}(t') \left[ \int_{w(t',h)}^t \frac{A}{2\sqrt{\xi-w}}d\xi \right]^2 dt'. \qquad (A8)$$

Moreover, eqn.A6 gives $h(t',w) = B\sqrt{w-t'}$ which implies $w(t',h) = \frac{h^2}{B^2} + t'$. Performing the integral in eqn.A8 we eventually obtain

$$\xi_c(h,t) = H\left(t - \frac{h^2}{B^2}\right)\pi \int_0^{t-\frac{h^2}{B^2}} \dot{N}(t')\left[A\sqrt{t - \frac{h^2}{B^2} - t'}\right]^2 dt'. \qquad (A9)$$

For site-saturation nucleation eqn.A9 gives

$$\xi_c(h,t) = H\left(t - \frac{h^2}{B^2}\right)\pi N_0 \int_0^{t-\frac{h^2}{B^2}}\left[A\sqrt{t - \frac{h^2}{B^2} - t'}\right]^2 \delta(t')dt', \qquad (A10)$$

that leads to eqn.A4.

*Cone-shaped nuclei.* In the case of right circular cones $p_r = \frac{\mu_1}{\sqrt{t}}$, $p_h = \frac{\mu_2}{\sqrt{t}}$ (with constant $\mu_i$) and $w(t',h) = (\frac{h}{2\mu_2} + \sqrt{t'})^2$. For site saturation nucleation eqn.A7 becomes ($t'_{mx} = \sqrt{t} - \frac{h}{2\mu_2}$)

$$\xi_c(h,t) = H(t'_{mx})\pi N_0 \int_0^{t'_{mx}} \delta(t')\left[\int_{(\frac{h}{2\mu_2}+\sqrt{t'})^2}^t \frac{\mu_1}{\sqrt{t}}\, d\xi\right]^2 dt' = 4\pi N_0 \mu_1^2 \left[\sqrt{t} - \frac{h}{2\mu_2}\right]^2 \quad (A11)$$

The volume of the deposit is equal to,

$$V(t) = \int_0^{h_{mx}}(1 - e^{-\xi_c(h,t)})\, dh = 2\mu_2\sqrt{t}\int_0^1(1 - e^{-4\pi N_0 \mu_1^2 t \xi^2})\, d\xi. \qquad (A12)$$

The current density is obtained from the time derivative of eqn.A12, $J(t) = zF\frac{\rho}{M}\frac{dV(t)}{dt}$, as

$$J(t) = \frac{zF\rho}{M}\frac{\mu_2}{\sqrt{t}}(1 - e^{-4\pi N_0\mu_1^2 t}).$$

(A13)

that is eqn.19.

**Figure captions**

Fig.1

a) Very early stage of the deposition. The left side of the figure shows a pictorial view of well separated hemispherical nuclei randomly distributed on the electrode surface. The arrows provide a schematic representation of the radial diffusion field close to nucleus surface. In this case, radial diffusion fields of different nuclei do not overlap. On the right side, equivalent planar system with a 1:1 correspondence between nuclei and hypothetical diffusion cylinders of radius $r_d$ and height $\bar{\delta}$. The height of the cylinder is equal to the Cottrell thickness of the diffusion layer. Arrows represent linear diffusion field at the base of the cylinder (planar diffusion zone). The current at the planar zone equals the growth current of the hemispherical nucleus (eqn.1a).

b-The left side of the figure refers to a later time of the growth. In this case, overlap among radial diffusion fields and impingement among nuclei both take place. In the equivalent planar system, displayed on the right side of the figure, such a situation implies overlapping diffusion cylinders.

c- "3D nucleation and growth" model. Schematic representation of the critical region for the generic point of the space, Q, located at height $h < h_{mx} = R$. The probability the point Q is not transformed by the new phase up to $t$, is linked to the area of the critical region through eqns.9a,b. The red disk is the critical region (or capture zone) for Q: nucleation events occurring within the disk are capable of transforming Q up to $t$. The hemisphere centered at N represents a nucleus just at the border of the critical region.

Fig.2- Kinetics of film growth according to the "3D nucleation and growth" (curve 3D) and "planar diffusion zones" models. The volume per unit area, $V$, has been normalized to $(\beta_0 t)^{1/2}$ and is reported as a function of reduced time, $S_{ex}$. For the planar model the kinetics are computed for v=2, v=5, v=10 and v=20 (blue curves). The inset shows the behavior of the normalized current density for v=20 and v =30.

Fig.3- Behavior of the reduced nucleus radius, $R/\zeta$ as a function of reduced time, $\tau$, for the model case of binary collisions. Single ($n$=1) and double ($n$=2) collisions are considered for v=20, v=40 and v=70. The parabolic growth law employed in the original formulation of the model is also shown as black solid line. The figure indicates that for a given time, displacement from parabolic law is larger the higher $n$ and v. The best fit of power law to the solution of the differential equation is displayed as dashed lines.

Fig.4- The function $\tilde{R}(t)$ (solid symbols), calculated using eqn.12, is reported for a) $v = 10$, b) $v = 70$ and c) $v = 100$. The best fit (red solid line) of $\tilde{R}(t)$ is obtained using the power law $\tilde{R} = aS_{ex}^k$. In the same figure the fraction of the electrode surface covered by islands is displayed (dotted line). The parabolic growth is also reported as blue solid line.

Fig.5- Behavior of the exponent ($\kappa$) and pre-factor ($a$) of the power law $\tilde{R} = aS_{ex}^k$ as a function of $\nu$. The reduced nucleus radius, $\tilde{R}$, has been obtained by solving eqn.12. The power exponent for the model case of binary collisions is also reported for $n$=1 (blue symbols) and $n$=2 (black symbols). The dashed horizontal lines marks the range of variation of $\kappa$: $\frac{1}{6} < \kappa(\nu) < \frac{1}{2}$.

Fig.6- Substrate coverage, $S$, as a function of reduced time for $\nu = 70$ (red curves), $\nu = 100$ (blue curves) and $\nu = 130$ (black curves) and aspect ratio $\alpha = 0.5$ (dotted lines) and $\alpha = 1$ (solid lines). In the same figure it is also displayed $S$ for parabolic growth ($\kappa$=1/2) (solid grey line). In the case of parabolic growth law $S$ is independent of $\nu$. 2D images of computer simulations of 3D KJMA-type phase transformations are also reported showing the aspect of the surface at different $S$.

Fig.7- The behavior of the mean thickness of the deposit ($\bar{h}$) and fraction of the substrate surface covered by islands ($S$) are reported in panels a) and b), respectively, as a function of reduced time ($S_{ex}$) and aspect ratio ($\alpha$) for $\nu$=70.

Fig.A1 Illustration of the "capture zone" method for ellipsoidal nucleus. $\overline{ON}$ is the radius of the capture zone for point Q. Nuclei located at distance $d > \overline{ON}$ are unable to transform Q up to time $t$. In the framework of the two-rate approach of ref.[6], the capture zone (red disk) is located at height $h$ and is related to the radial growth rate of the nuclei, $p_r$.

**Figures**

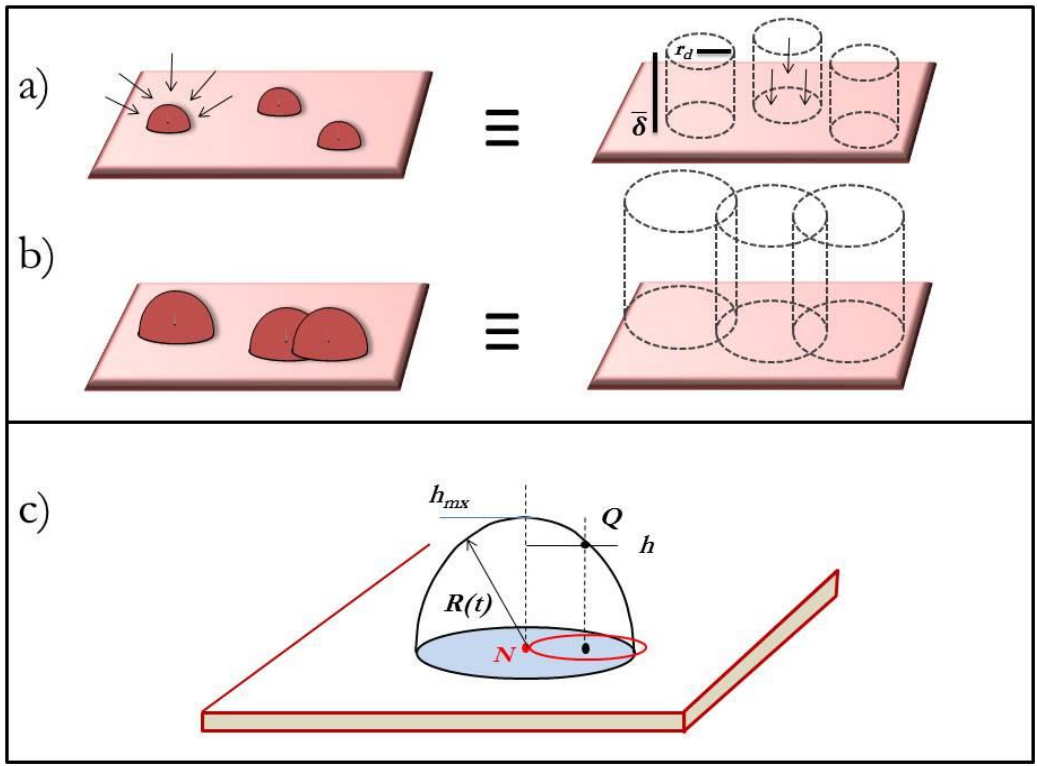

Fig.1

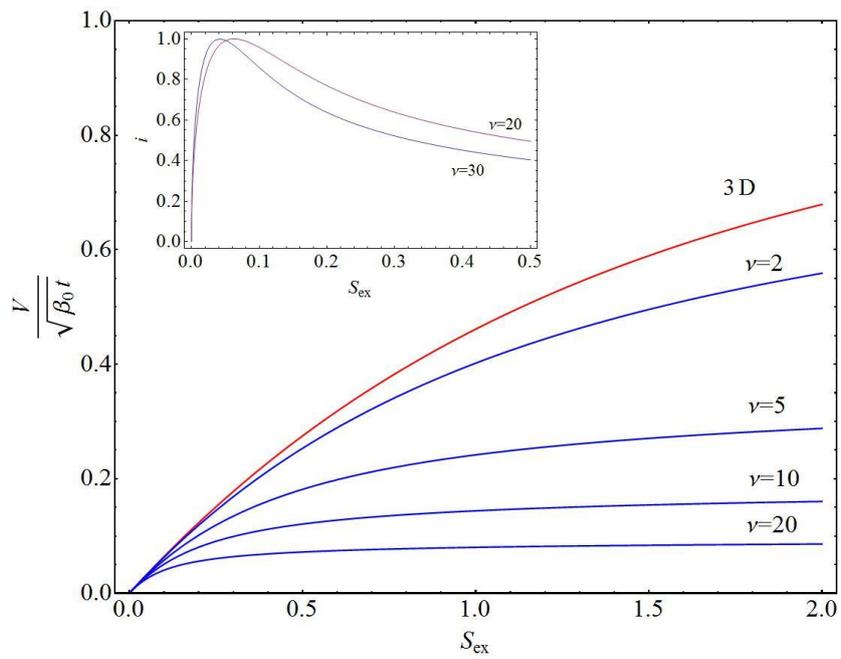

Fig.2

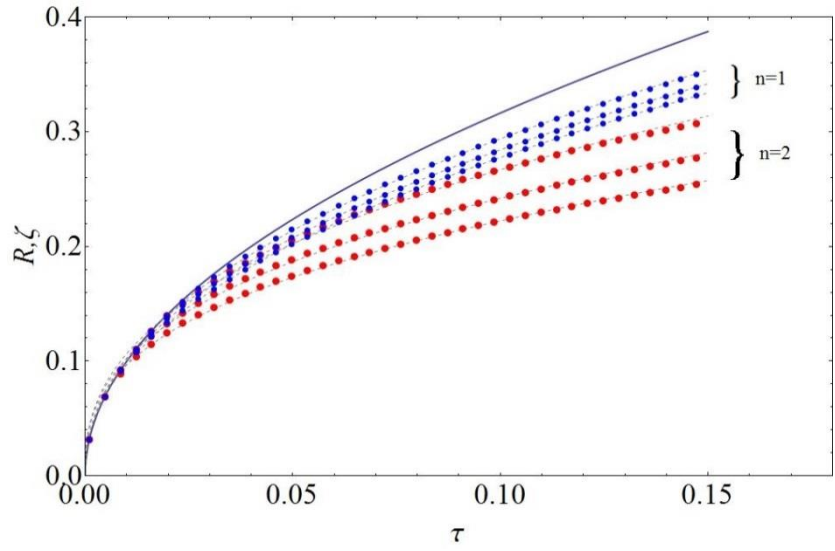

Fig.3

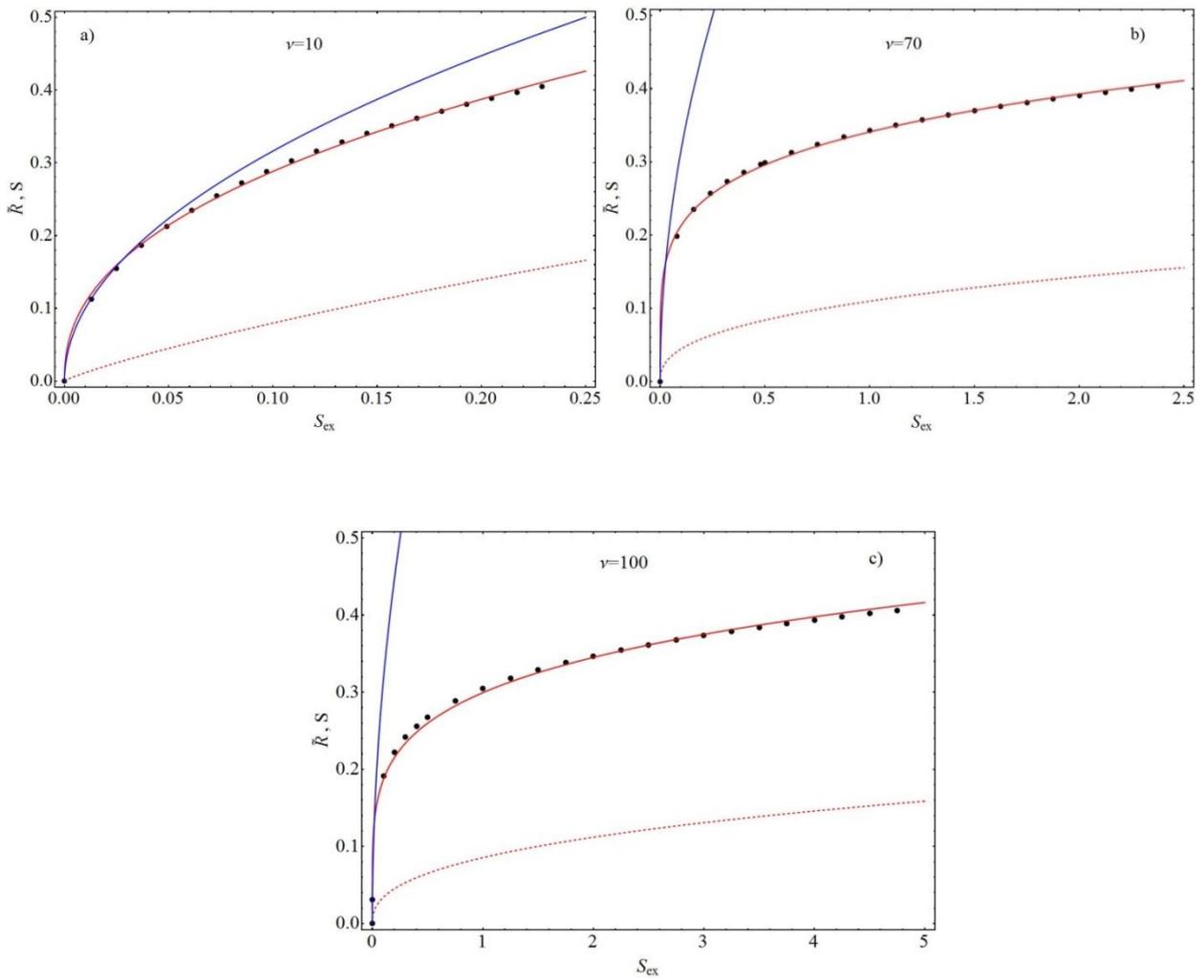

Fig.4

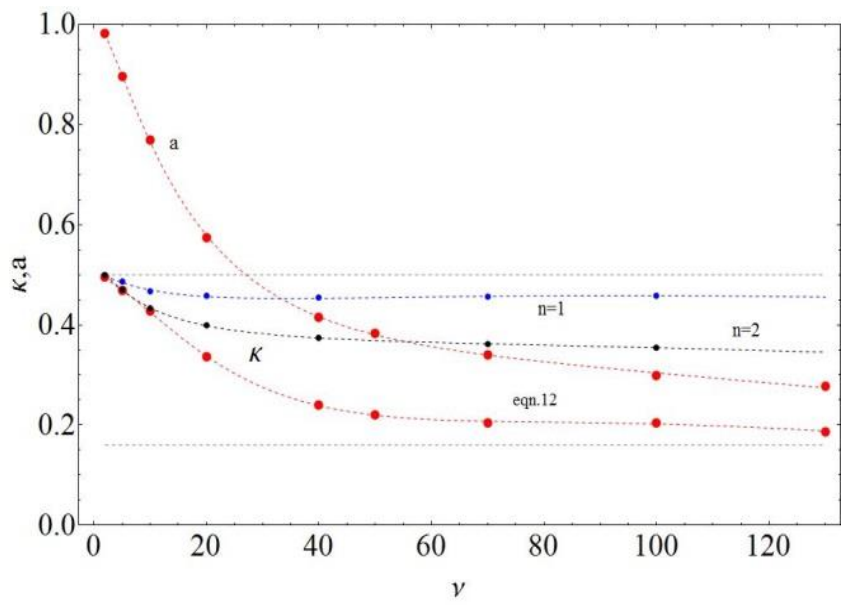

Fig.5

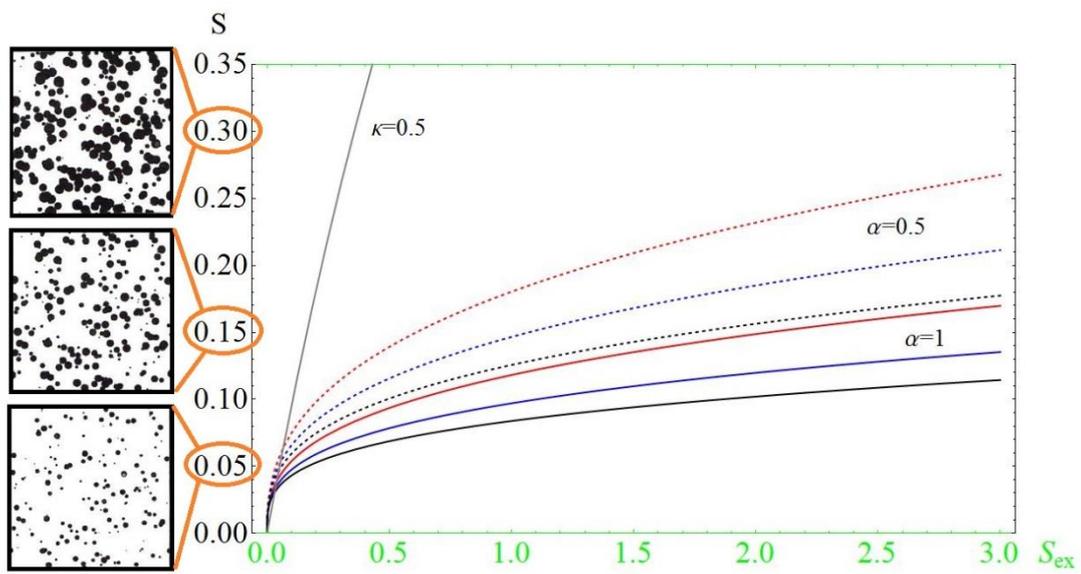

Fig.6

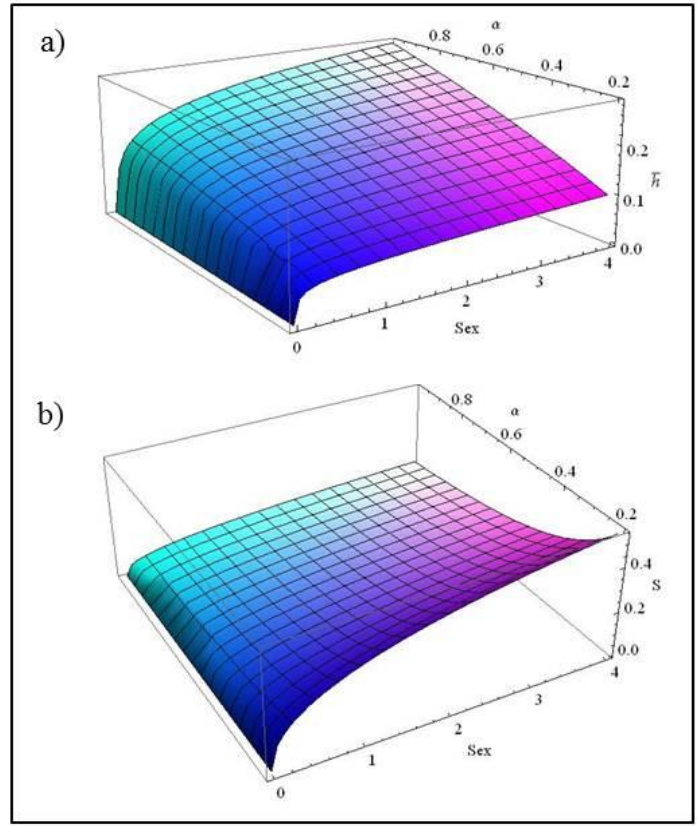

Fig.7

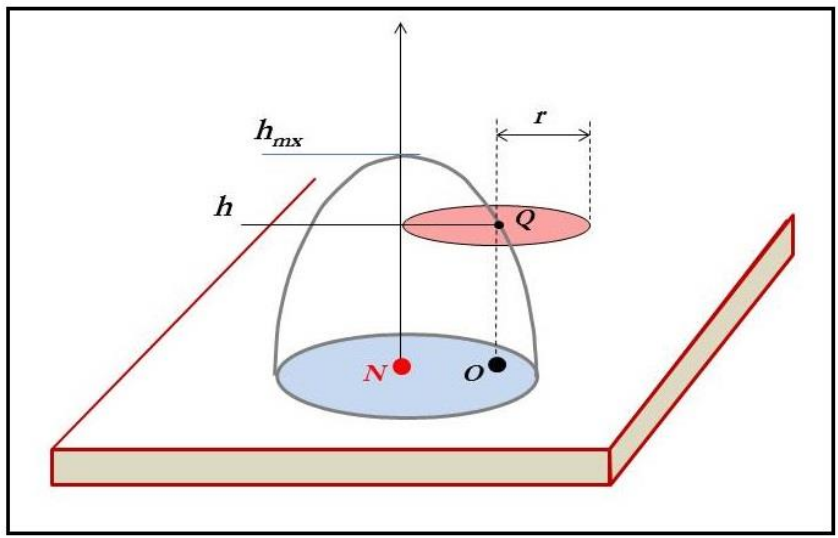

Fig.A1